# The Berlekamp-Massey Algorithm and the Euclidean Algorithm: a Closer Link

Maria Bras-Amorós*    Michael E. O'Sullivan

August 15, 2009

**Abstract**

The two primary decoding algorithms for Reed-Solomon codes are the Berlekamp-Massey algorithm and the Sugiyama et al. adaptation of the Euclidean algorithm, both designed to solve a key equation. In this article an alternative version of the key equation and a new way to use the Euclidean algorithm to solve it are presented, which yield the Berlekamp-Massey algorithm. This results in a new, simpler, and compacter presentation of the Berlekamp-Massey algorithm.

## 1 Introduction

For correcting Reed-Solomon codes the primary tools are the so-called locator and evaluator polynomials. Once we know them, the error positions are determined by the inverses of the roots of the locator polynomial and the error values can be computed by a formula due to Forney [3] which uses the evaluator and the derivative of the locator evaluated at the inverses of the error positions.

Berlekamp presented in [1] a key equation determining the decoding polynomials for primal Reed-Solomon codes. The Berlekamp-Massey algorithm, which unites Massey's perspective in [6] with Berlekamp's work, is the most prominent algorithm for decoding Reed-Solomon codes. It is based on linear feedback shift registers. Sugiyama et al. introduced in [8] an alternative algorithm for solving the key equation based on the Euclidean algorithm for computing the greatest common divisor of two polynomials and the coefficients of the Bézout identity. The Berlekamp-Massey algorithm is widely accepted to have better performance than the Sugiyama algorithm, although its formulation is perhaps more difficult to understand. The connections between the two algorithms were analyzed in [2, 4].

*This work was partly supported by the Catalan Government through a grant 2005 SGR 00446 and by the Spanish Ministry of Education through projects TSI2007-65406-C03-01 E-AEGIS and CONSOLIDER CSD2007-00004 ARES.

In this work we take the perspective of dual Reed Solomon codes. In Section 2 we present a key equation for dual Reed-Solomon codes and in Section 4 we introduce a Euclidean-based algorithm for solving it, following Sugiyama's idea. While the standard key equation for primal Reed-Solomon codes states that a linear combination of the decoding polynomials is a multiple of a certain power of $x$, in the key equation presented here the linear combination has bounded degree. Another important difference is that in the Sugiyama algorithm the locator and evaluator polynomials play the role of one of the Bézout's coefficients and the remainder respectively, while in the Euclidean algorithm presented here the locator and evaluator polynomials play the role of the two coefficients of the Bézout identity.

The decoding polynomials are now slightly different and in a sense more natural, since the error positions are given by the roots themselves of the locator polynomial rather than their inverses and the error values are obtained by evaluating the evaluator polynomial and the derivative of the locator polynomial at the error positions rather than evaluating them at the inverses of the error positions. In addition, the equivalent of the Forney formula does not have the minus sign.

In Section 5 we show that the intermediate remainders computed in the Euclidean algorithm are not necessary to find the Bézout coefficients. This leads to a proof of the fact that the new Euclidean-based algorithm is the Berlekamp-Massey algorithm.

## 2 Key equations for decoding Reed-Solomon codes revisited

In this section we will first establish the notions and notations related to Reed-Solomon codes that we will use in the present work. A general reference is [7]. We will present the usual decoding polynomials for primal codes, and analogous —but different—decoding polynomials for dual codes. We will see that each set of polynomials satisfies a key equation, which can be written in the form of intermediate Bézout identities. In the primal case the key equation is Berlekamp's key equation and the adaptation of the extended Euclidean algorithm for solving it is the algorithm of Sugiyama et al. In the dual case the key equation is new and the discussion of the extended Euclidean algorithm for solving it is the aim of the present work. It turns out that the definitions we use for dual codes, are more natural in the context of general codes from curves; see *e*.g. [5].

**Settings and notations**

Let $\mathbb{F}$ be a finite field of size $q$ and let $\alpha$ be a primitive element in $\mathbb{F}$. Let $n = q - 1$. We identify the vector $u = (u_0, \ldots, u_{n-1})$ with the polynomial $u(x) = u_0 + \cdots + u_{n-1}x^{n-1}$ and denote $u(a)$ the evaluation of $u(x)$ at $a$. Classically the Reed-Solomon code $C^*(k)$ of dimension $k$ is defined as the cyclic code with generator polynomial

$$(x - \alpha)(x - \alpha^2) \cdots (x - \alpha^{n-k}),$$

Its generator matrix is

$$G^*(k) = \begin{pmatrix} 1 & 1 & 1 & \ldots & 1 \\ 1 & \alpha & \alpha^2 & \ldots & \alpha^{n-1} \\ 1 & \alpha^2 & \alpha^4 & \ldots & \alpha^{2(n-1)} \\ \vdots & \vdots & \vdots & \vdots & \vdots \\ 1 & \alpha^{k-1} & \alpha^{(k-1)2} & \ldots & \alpha^{(k-1)(n-1)} \end{pmatrix}$$

and its parity check matrix is

$$H^*(k) = \begin{pmatrix} 1 & \alpha & \alpha^2 & \ldots & \alpha^{n-1} \\ 1 & \alpha^2 & \alpha^4 & \ldots & \alpha^{2(n-1)} \\ 1 & \alpha^3 & \alpha^6 & \ldots & \alpha^{3(n-1)} \\ \vdots & \vdots & \vdots & \vdots & \vdots \\ 1 & \alpha^{n-k} & \alpha^{(n-k)2} & \ldots & \alpha^{(n-k)(n-1)} \end{pmatrix}.$$

The dual Reed-Solomon code $C(k)$ of dimension $k$ is the cyclic code with generator polynomial

$$(x - \alpha^{-(k+1)})(x - \alpha^{-(k+2)}) \cdots (x - \alpha^{-(n-1)})(x - 1).$$

Its generator matrix is

$$G(k) = \begin{pmatrix} 1 & \alpha & \alpha^2 & \ldots & \alpha^{n-1} \\ 1 & \alpha^2 & \alpha^4 & \ldots & \alpha^{2(n-1)} \\ 1 & \alpha^3 & \alpha^6 & \ldots & \alpha^{3(n-1)} \\ \vdots & \vdots & \vdots & \vdots & \vdots \\ 1 & \alpha^k & \alpha^{2k} & \ldots & \alpha^{k(n-1)} \end{pmatrix}$$

and its parity check matrix is

$$H(k) = \begin{pmatrix} 1 & 1 & 1 & \ldots & 1 \\ 1 & \alpha & \alpha^2 & \ldots & \alpha^{n-1} \\ 1 & \alpha^2 & \alpha^4 & \ldots & \alpha^{2(n-1)} \\ \vdots & \vdots & \vdots & \vdots & \vdots \\ 1 & \alpha^{n-k-1} & \alpha^{(n-k-1)2} & \ldots & \alpha^{(n-k-1)(n-1)} \end{pmatrix}.$$

To emphasize the difference between Reed-Solomon codes and their duals we will call the former ones primal Reed-Solomon codes. Both codes have minimum distance $d = n - k + 1$. Furthermore, $C(k)^\perp = C^*(n - k)$. There is a natural bijection from $\mathbb{F}^n$ to itself which we denote by $c \mapsto c^*$. It takes $C(k)$ to $C^*(k)$. The codeword $c^*$ can be defined either as $iG^*(k) \in C^*(k)$ where $i$ is the information vector of dimension $k$ such that $c = iG(k) \in C(k)$ or componentwise as $c^* = (c_0, \alpha^{-1}c_1, \alpha^{-2}c_2, \ldots, \alpha c_{n-1})$ where $c = (c_0, c_1, \ldots, c_{n-1})$. Then,
$$c = (c_0^*, \alpha c_1^*, \alpha^2 c_2^*, \ldots, \alpha^{n-1} c_{n-1}^*). \tag{1}$$
In particular, $c(\alpha^i) = c_0^* + \alpha c_1^* \alpha^i + \alpha^2 c_2^* \alpha^{2i} + \cdots + \alpha^{n-1} c_{n-1}^* \alpha^{(n-1)i} = c^*(\alpha^{i+1})$.

A decoding algorithm for a primal Reed-Solomon code may be used to decode a dual Reed-Solomon code by first applying the bijection $*$ to the received vector $u$. If $u$ differs from a codeword $c \in C(k)$ by an error vector $e$ of weight $t$, then $u^*$ differs from the codeword $c^* \in C^*(k)$ by the error vector $e^*$ of weight $t$. If the primal Reed-Solomon decoding algorithm can decode $u^*$ to obtain $c^*$ and $e^*$ then, transforming by the inverse of $*$ we may obtain $c$ and $e$. Conversely, a decoding algorithm for a dual Reed-Solomon code may be used to decode a primal Reed-Solomon code by applying the inverse of $*$, decoding, and then applying $*$.

**Key polynomials**

Suppose that a word $c^* \in C^*(k)$ is transmitted and that an error $e^*$ occurred, so that $u^* = c^* + e^*$ is received. Define the *error locator polynomial* $\Lambda^*$ and the *error evaluator polynomial* $\Omega^*$ as
$$\Lambda^* = \prod_{i: e_i^* \neq 0} (1 - \alpha^i x), \quad \Omega^* = \sum_{i: e_i^* \neq 0} e_i^* \alpha^i \prod_{j: e_j^* \neq 0, j \neq i} (1 - \alpha^i x).$$

If $\Lambda^*$ and $\Omega^*$ are known, the error positions can be identified as the indices $i$ such that
$$\Lambda^*(\alpha^{-i}) = 0$$
and the error values can be computed by the Forney formula [3]
$$e_i^* = -\frac{\Omega^*(\alpha^{-i})}{\Lambda^{*\prime}(\alpha^{-i})}.$$

Analogously, for dual Reed-Solomon codes, suppose that a word $c \in C(k)$ is transmitted and that an error $e$ occurred, so that $u = c + e$ is received. In this case define the *error locator polynomial* $\Lambda$ and the *error evaluator polynomial* $\Omega$ as
$$\Lambda = \prod_{i: e_i \neq 0} (x - \alpha^i), \quad \Omega = \sum_{i: e_i \neq 0} e_i \prod_{j: e_j \neq 0, j \neq i} (x - \alpha^i).$$

Now, if $\Lambda$ and $\Omega$ are known, the error positions can be identified as the indices $i$ such that
$$\Lambda(\alpha^i) = 0$$
and the Forney formula for error values is somewhat simpler,
$$e_i = \frac{\Omega(\alpha^i)}{\Lambda'(\alpha^i)}.$$

It is easy to check that when the error vectors $e$ and $e^*$ satisfy the relationship (1), then the polynomials $\Lambda$, $\Omega$, associated to the error $e$ are related to the polynomials $\Lambda^*$, $\Omega^*$, associated to the error $e^*$ as follows:

$$\Lambda = x^t \Lambda^*(1/x), \quad \Omega = x^{t-1}\Omega^*(1/x), \qquad (2)$$

where $t = |e| = |e^*|$.

## Key equations

The *syndrome polynomial* is defined for primal codes as

$$S^* = e^*(\alpha) + e^*(\alpha^2)x + e^*(\alpha^3)x^2 + \cdots + e^*(\alpha^n)x^{n-1}.$$

For dual codes define

$$S = e(\alpha^{n-1}) + e(\alpha^{n-2})x + \cdots + e(\alpha)x^{n-2} + e(1)x^{n-1}.$$

Notice that while the general term of $S^*$ is $e^*(\alpha^{i+1})x^i$, the general term of $S$ is $e(\alpha^{n-1-i})x^i$. If the error vectors $e$ and $e^*$ are related by ((1)), then

$$S = x^{n-1}S^*(1/x). \qquad (3)$$

It is easily verified that
$$\Lambda S = (x^n - 1)\Omega \qquad (4)$$

and, by (2) and (3),
$$\Lambda^* S^* = (-x^n + 1)\Omega^*.$$

For primal codes, from the received word we only know $e^*(\alpha) = u^*(\alpha), \ldots, e^*(\alpha^{n-k}) = u^*(\alpha^{n-k})$. This is why we use the *truncated syndrome polynomial*

$$\bar{S}^* = e^*(\alpha) + e^*(\alpha^2)x + \cdots + e^*(\alpha^{n-k})x^{n-k-1}.$$

Notice that $\Lambda^*\bar{S}^* + (x^n - 1)\Omega^* = \Lambda^*(\bar{S}^* - S^*)$ has only terms of order $n - k$ or larger and this implies

$$\boxed{\Lambda^*\bar{S}^* = \Omega^* \mod x^{n-k}}$$

This is the key equation introduced by Berlekamp [1]. Massey [6] gave an algorithm for solving the key equation using linear feedback shift registers. Sugiyama et al. [8] recognized that the extended Euclidean algorithm could also be adapted to solve this kind of equation for $\Lambda^*$ and $\Omega^*$, starting with $r_{-2} = x^{n-k}$ and $r_{-1} = \bar{S}^*$. Their method is often referred to as the Euclidean decoding algorithm for Reed Solomon codes.

Analogously, for dual codes, from the received word we only know $e(1) = u(1), \ldots, e(\alpha^{n-k-1}) = u(\alpha^{n-k-1})$. In this case we use the truncated syndrome polynomial

$$\bar{S} = e(\alpha^{n-k-1})x^k + e(\alpha^{n-k-2})x^{k+1} + \cdots + e(1)x^{n-1}.$$

Observe that the polynomial $\Lambda \bar{S} - (x^n - 1)\Omega = \Lambda(\bar{S} - S)$ has degree less than $d/2 + k - 1 = n - d/2$. That is

$$\deg(\Lambda \bar{S} - (x^n - 1)\Omega) < n - d/2 \tag{5}$$

The aim of the present work is to deal with this alternative key equation for $\Lambda$ and $\Omega$, solving it by the extended Euclidean algorithm starting with $r_{-2} = x^n - 1$ and $r_{-1} = \bar{S}$.

## 3  Extended Euclidean algorithm revisited

Let $\mathbb{F}$ be a field. It is well known that given two polynomials $a, b \in \mathbb{F}[x]$ there exist two polynomials $f, g \in \mathbb{F}[x]$ such that

$$fb + ga = \gcd(a, b).$$

This equality is commonly known as Bézout's identity. We refer to $f$ and $g$ as the Bézout coefficients. The extended Euclidean algorithm computes not only $\gcd(a, b)$ but also the Bézout coefficients. For all $i \geqslant 0$, the algorithm produces $f_i, g_i$ and $r_i$ such that

$$f_i b + g_i a = r_i. \tag{6}$$

and for some $n$, $r_n = \gcd(a, b)$. We will refer to the equalities (6) as Bézout's intermediate identities and the coefficients $f_i$ and $g_i$ as the intermediate Bézout coefficients. The extended Euclidean algorithm is initialized with $r_{-2} = a$, $r_{-1} = b$ and $f_{-1} = 1$, $g_{-1} = 0$, $f_{-2} = 0$, $g_{-2} = 1$. For $i \geqslant 0$, let the Euclidean division of $r_{i-2}$ by $r_{i-1}$ be $r_{i-2} = q_i r_{i-1} + r_i$, and let $f_i = f_{i-2} - q_i f_{i-1}$ and $g_i = g_{i-2} - q_i g_{i-1}$. The algorithm may then be expressed in matrix form as

$$\begin{pmatrix} r_i & f_i & g_i \\ r_{i-1} & f_{i-1} & g_{i-1} \end{pmatrix} = \begin{pmatrix} -q_i & 1 \\ 1 & 0 \end{pmatrix} \begin{pmatrix} r_{i-1} & f_{i-1} & g_{i-1} \\ r_{i-2} & f_{i-2} & g_{i-2} \end{pmatrix} \tag{7}$$

together with the initial condition

$$\begin{pmatrix} r_{-1} & f_{-1} & g_{-1} \\ r_{-2} & f_{-2} & g_{-2} \end{pmatrix} = \begin{pmatrix} b & 1 & 0 \\ a & 0 & 1 \end{pmatrix}.$$

At each iteration we have the matrix equation $\begin{pmatrix} f_i & g_i \\ f_{i-1} & g_{i-1} \end{pmatrix} \begin{pmatrix} r_{-1} \\ r_{-2} \end{pmatrix} = \begin{pmatrix} r_i \\ r_{i-1} \end{pmatrix}$.

We will make a series of alterations of the algorithm: first ensuring that certain polynomials are monic throughout the algorithm; second breaking the algorithm into steps for each individual coefficient of the quotient polynomials.

Assume that $a$ is monic. Consider the algorithm

**Initialize:**

$$\begin{pmatrix} r_{-1} & f_{-1} & g_{-1} \\ \tilde{r}_{-1} & \tilde{f}_{-1} & \tilde{g}_{-1} \end{pmatrix} = \begin{pmatrix} b & 1 & 0 \\ a & 0 & 1 \end{pmatrix}$$

**while** $r_i \neq 0$,

$\mu_i = \mathbf{LC}(r_i)$
$q_i = \mathbf{Quotient}(\mu_i \tilde{r}_i, r_i)$

$$\begin{pmatrix} r_{i+1} & f_{i+1} & g_{i+1} \\ \tilde{r}_{i+1} & \tilde{f}_{i+1} & \tilde{g}_{i+1} \end{pmatrix} = \begin{pmatrix} q_i & -\mu_i \\ \mu_i^{-1} & 0 \end{pmatrix} \begin{pmatrix} r_i & f_i & g_i \\ \tilde{r}_i & \tilde{f}_i & \tilde{g}_i \end{pmatrix}$$

**end while**

**Return** $\tilde{r}_i, \tilde{f}_i, \tilde{g}_i$

It is clear that $\gcd(r_i, \tilde{r}_i) = \gcd(a, b)$ for all $i$. Furthermore, $\tilde{r}_i$ is monic by construction and $q_i$ is also monic. Clearly, $\deg(\tilde{f}_{i+1}) = \deg(f_i)$, and one can see using induction that $\deg(f_i)$ is strictly increasing and that the $f_i$ are also monic.

Now consider the division of a monic polynomial $a$ by $b \neq 0$. Let $\mu = \mathrm{LC}(b)$, let $d = \deg(a) - \deg(b)$ and let $q(x) = \mathbf{Quotient}(\mu a, b) = q_0 + q_1 x + \cdots + q_{d-1} x^{d-1} + x^d$. We have the factorization $\begin{pmatrix} q(x) & -\mu \\ \mu^{-1} & 0 \end{pmatrix} = \begin{pmatrix} 1 & \mu q_0 \\ 0 & 1 \end{pmatrix} \begin{pmatrix} 1 & \mu q_1 x \\ 0 & 1 \end{pmatrix} \cdots \begin{pmatrix} 1 & \mu q_{d-1} \\ 0 & 1 \end{pmatrix}$

Let $\begin{pmatrix} R_0 \\ \tilde{R}_0 \end{pmatrix} = \begin{pmatrix} b \\ a \end{pmatrix}$, and $\begin{pmatrix} R_1 \\ \tilde{R}_1 \end{pmatrix} = \begin{pmatrix} x^d & -\mu \\ \mu^{-1} & 0 \end{pmatrix} \begin{pmatrix} b \\ a \end{pmatrix} = \begin{pmatrix} x^d b - \mu a \\ \mu^{-1} b \end{pmatrix}$, and inductively define,

$$\begin{pmatrix} R_j \\ \tilde{R}_j \end{pmatrix} = \begin{pmatrix} 1 & \mu q_{d-j+1} x^{d-j+1} \\ 0 & 1 \end{pmatrix} \begin{pmatrix} R_{j-1} \\ \tilde{R}_{j-1} \end{pmatrix}$$
$$= \begin{pmatrix} x^d + q_{d-1} x^{d-1} + \cdots + q_{d-j+1} x^{d-j+1} & -\mu \\ \mu^{-1} & 0 \end{pmatrix} \begin{pmatrix} b \\ a \end{pmatrix}$$

Note that for $j > 0$, $\tilde{R}_j = \mu^{-1} b$ is monic and that $\deg(R_j) \leqslant \deg(b) + d - j = \deg(a) - j$ with $\mu q_{d-j} = -\mathbf{Coefficient}(R_j, \deg(a) - j)$. We may think of $\deg(a) - j$ as a bound on the degree of the partial remainders $R_j$ that decreases as $j$ increases. As $j$ goes from 0 to 1 we have $\deg(\tilde{R}_1) = \deg(R_0)$ and $\deg(R_1) \leqslant \deg(\tilde{R}_0) - 1$.

We now alter the variant of the Euclidean algorithm given above by factoring all of the matrices $\begin{pmatrix} q_i(x) & -\mu_i \\ \mu_i^{-1} & 0 \end{pmatrix}$ and including an iteration for each factor. We assume $a$ is monic. We include two counters, $\tilde{d}_j$ is the degree of $\tilde{R}_j$, and $d_j$ is an upper bound for the degree of $R_j$.

**Initialize:**

$d_0 = \mathbf{deg}(b)$
$\tilde{d}_0 = \mathbf{deg}(a)$
$\begin{pmatrix} R_0 & F_0 & G_0 \\ \tilde{R}_0 & \tilde{F}_0 & \tilde{G}_0 \end{pmatrix} = \begin{pmatrix} b & 1 & 0 \\ a & 0 & 1 \end{pmatrix}$

**while** $d_i \geqslant 0$,

    $\mu_i = \textbf{Coefficient}(R_i, d_i)$
    $p = d_i - \tilde{d}_i$
    **if** $p \geqslant 0$ **or** $\mu_i = 0$ **then**

$$\begin{pmatrix} R_{i+1} & F_{i+1} & G_{i+1} \\ \tilde{R}_{i+1} & \tilde{F}_{i+1} & \tilde{G}_{i+1} \end{pmatrix} = \begin{pmatrix} 1 & -\mu_i x^p \\ 0 & 1 \end{pmatrix} \begin{pmatrix} R_i & F_i & G_i \\ \tilde{R}_i & \tilde{F}_i & \tilde{G}_i \end{pmatrix}$$

$$d_{i+1} = d_i - 1$$
$$\tilde{d}_{i+1} = \tilde{d}_i$$

    **else**

$$\begin{pmatrix} R_{i+1} & F_{i+1} & G_{i+1} \\ \tilde{R}_{i+1} & \tilde{F}_{i+1} & \tilde{G}_{i+1} \end{pmatrix} = \begin{pmatrix} x^{-p} & -\mu_i \\ 1/\mu_i & 0 \end{pmatrix} \begin{pmatrix} R_i & F_i & G_i \\ \tilde{R}_i & \tilde{F}_i & \tilde{G}_i \end{pmatrix}$$

$$d_{i+1} = \tilde{d}_i - 1$$
$$\tilde{d}_{i+1} = d_i$$

    **end if**

**end while**

**Return** $\tilde{R}_i, \tilde{F}_i, \tilde{G}_i$

**Proposition 1.** *In the algorithm above, for any $i$,*

1. $\begin{pmatrix} F_i & G_i \\ \tilde{F}_i & \tilde{G}_i \end{pmatrix} \begin{pmatrix} b \\ a \end{pmatrix} = \begin{pmatrix} R_i \\ \tilde{R}_i \end{pmatrix}$.

2. *$\tilde{R}_i$ is monic of degree $\tilde{d}_i$;*

3. *$R_i$ has degree at most $d_i$;*

4. *$d_i + \tilde{d}_i = \deg(a) + \deg(b) - i$;*

5. *$F_i \tilde{G}_i - \tilde{F}_i G_i = 1$.*

6. *$F_i$ is monic and*

$$\deg(a) = \deg(F_i) + \tilde{d}_i > \deg(\tilde{F}_i) + d_i$$

*Furthermore, at termination, $\tilde{R}_i = \tilde{F}_i b + \tilde{G}_i a$ is the monic gcd of $a$ and $b$ and $F_i b + G_i a = 0$ with $F_i$ monic.*

*Proof.* These results are proven by induction, with the base step an easy verification. We must check the induction step for each of the two update formulas—the first with $p > 0$ or $\mu_i = 0$ and the second $p < 0$ and $\mu_i \neq 0$.

Item (1) is immediate to prove using the fact that both $\begin{pmatrix} F_i & G_i \\ \tilde{F}_i & \tilde{G}_i \end{pmatrix}$ and $\begin{pmatrix} R_i \\ \tilde{R}_i \end{pmatrix}$ are updated by multiplying by the same matrix.

For item (2), note that the first update doesn't change $\tilde{R}$ and $\tilde{d}$, so the result is immediate. With the second update $\tilde{R}_{i+1} = R_i / \operatorname{LC}(R_i)$, so it is monic of degree $d_i = \tilde{d}_{i+1}$.

For item (3), and the first update, the coefficient of $x^{d_i}$ in both $R_i$ and $\mu_i x^p \tilde{R}_i$ is $\mu_i$ so $\deg(R_{i+1}) < d_i$ as desired. In the second update, the coefficient of $x^{\tilde{d}_i}$ in both $x^{-p} R_i$ and $\mu_i \tilde{R}_i$ is $\mu_i$ so $\deg(R_{i+1}) < \tilde{d}_i$ as desired.

Item (4) is immediately verified. Item (5) follows from the inductive hypothesis by taking determinants in the update equation. For example, with the second update,

$$\begin{pmatrix} F_{i+1} & G_{i+1} \\ \tilde{F}_{i+1} & \tilde{G}_{i+1} \end{pmatrix} = \begin{pmatrix} x^{-p} & -\mu_i \\ \mu_i^{-1} & 0 \end{pmatrix} \begin{pmatrix} F_i & G_i \\ \tilde{F}_i & \tilde{G}_i \end{pmatrix}$$

For item (6), consider the first update: $F_{i+1} = F_i - \mu_i x^{d_i - \tilde{d}_i} \tilde{F}_i$. By the induction hypothesis, $\deg(F_i) > \deg(\tilde{F}_i) + d_i - \tilde{d}_i$, so $\deg(F_{i+1}) = \deg(F_i)$ and $F_{i+1}$ is monic, since $F_i$ is. Similarly, with the second update, $\deg(F_{i+1}) = \deg(x^{\tilde{d}_i - d_i} F_i) > \deg(\tilde{F}_i)$, so $F_{i+1}$ is monic. As with item (5), using the observation that the determinants of the update matrices are 1, we can prove that $F_i \tilde{R}_i - \tilde{F}_i R_i = -a$. Taking degrees gives the statement of item (6).

It is easy to see that $\gcd(R_i, \tilde{R}_i) = \gcd(a, b)$, for all $i$. At termination, we have $R_i = 0$ so $\gcd(a, b) = \tilde{R}_i$. This gives the final statement of the proposition. □

## 4 Euclidean algorithm for the new key equation

Suppose $u = c + e$ is the received word, with $c$ belonging to the dual Reed-Solomon code $C(k)$ and the error $e$ having weight $t$. Let $\Lambda$, $\Omega$, $S$ and $\bar{S}$ be the polynomials associated to $e$. We show that the extended Euclidean algorithm may be used to solve the alternative key equation (5).

**Lemma 2.** *Let $f \in \mathbb{F}[x]$ and $\alpha \in \mathbb{F}^*$ be such that $\deg(f) < n$ and $f(x)\frac{x^n - 1}{x - \alpha}$ has no term of degree $n - 1$. Then $f(\alpha) = 0$.*

*Proof.* Dividing $f$ by $x - \alpha$, there exists $g \in \mathbb{F}[x]$ with $\deg(g) < n - 1$ such that $f(x) = f(\alpha) + (x - \alpha)g(x)$. Then $f(x)\frac{x^n - 1}{x - \alpha} = f(\alpha)\frac{x^n - 1}{x - \alpha} + g(x)(x^n - 1)$. While $g(x)(x^n - 1)$ has no terms with degrees between $\deg(g) + 1$ ($\leqslant n - 1$) and $n - 1$, $f(\alpha)\frac{x^n - 1}{x - \alpha}$ has coefficient of degree $n - 1$ equal to $f(\alpha)$. Therefore, if the term of $f(x)\frac{x^n - 1}{x - \alpha}$ of degree $n - 1$ is zero, then $f(\alpha) = 0$. □

**Lemma 3.** *If $\deg(f) \leqslant n - t$ and $fS$ has no terms of degrees $n - t, \ldots, n - 1$, then $f$ is a multiple of $\Lambda$.*

*Proof.* Suppose $fS$ has no terms of degrees $n - t, \ldots, n - 1$. Suppose $e_j \neq 0$ and let
$$g(x) = \prod_{\substack{k : e_k \neq 0 \\ k \neq j}} (x - \alpha_k).$$

Note that $\deg(g) = t-1$ and so $fgS$ has no term of degree $n-1$. Now,

$$\begin{aligned}
fgS &= f(x)g(x)\frac{\Omega(x)(x^n-1)}{\Lambda(x)} \\
&= \sum_{k:e_k\neq 0} e_k f(x)g(x)\frac{x^n-1}{x-\alpha_k} \\
&= e_j f(x)g(x)\frac{x^n-1}{x-\alpha_j} + \sum_{\substack{k:e_k\neq 0 \\ k\neq j}} e_k f(x)(x^n-1)\frac{g(x)}{x-\alpha_k}.
\end{aligned}$$

Since both $fgS$ and the right hand term in the previous sum have no component of degree $n-1$, neither does $f(x)g(x)\frac{x^n-1}{x-\alpha_j}$. By the previous lemma, $x-\alpha_j$ must divide $f$. Since $j$ was chosen arbitrarily such that $e_j \neq 0$, we conclude that $\Lambda$ must divide $f$. □

The next lemma characterizes the decoding polynomials by means of the alternative key equation, the polynomial degrees, and their coprimality.

**Lemma 4.** *Suppose that $t$ errors occurred with $t < d/2$. Then $\Lambda$ and $\Omega$ are the unique polynomials $\lambda, \omega$ satisfying the following properties.*

1. $\deg(\lambda \bar{S} - \omega(x^n - 1)) < n - d/2$
2. $\deg(\lambda) \leqslant d/2$
3. $\lambda, \omega$ are coprime.
4. $\lambda$ is monic

*Proof.* We have shown that $\Lambda$ and $\Omega$ satisfy item 1, and items 2, 3, 4 are easily verified.

Suppose that $\lambda, \omega$ satisfy items 1, 2. We will show that $\lambda S$ has no terms in degrees $n-t, \ldots, n-1$. So, by Lemma 3, $\lambda$ is a multiple of $\Lambda$. Indeed, write

$$\lambda S = (\lambda \bar{S} - \omega(x^n - 1)) + \lambda(S - \bar{S}) + \omega(x^n - 1).$$

By 1, the first term has degree less than $n - d/2 < n - t$. By 2, $\deg(\lambda(S - \bar{S})) \leqslant d/2 + k - 1 = n - d/2 < n - t$. By 1, $\deg(\omega) < \deg(\lambda)$ and by 2, $\deg(\omega) < d/2 \leqslant n - d/2 < n - t$. So, $\omega(x^n - 1)$ has no terms in degrees $n - t, \ldots, n - 1$.

Suppose now that $\lambda = g\Lambda$ for some $g \in \mathbb{F}[x]$. Then

$$\begin{aligned}
\lambda \bar{S} - \omega(x^n - 1) &= -\lambda(S - \bar{S}) + \lambda S - \omega(x^n - 1) \\
&= -\lambda(S - \bar{S}) + g\Lambda S - \omega(x^n - 1) \\
&= -\lambda(S - \bar{S}) + g\Omega(x^n - 1) - \omega(x^n - 1) \\
&= -\lambda(S - \bar{S}) + (g\Omega - \omega)(x^n - 1).
\end{aligned}$$

By 1 $\deg(\lambda \bar{S} - \omega(x^n - 1)) < n - d/2$ and by 2, $\deg(\lambda(S - \bar{S})) \leqslant n - d/2$. As a consequence, $\omega = g\Omega$. Now items 3, 4 imply $g = 1$ and so $\lambda = \Lambda$ and $\omega = \Omega$. □

Consider the algorithm in Section 3 with $a = x^n - 1$ and $b = \bar{S}$, truncated when $d_i < n - d/2$, and returning $F_i, G_i$ instead of $\tilde{F}_i, \tilde{G}_i$:

**Initialize:**

$$d_0 = \mathbf{deg}(\bar{S})$$
$$\tilde{d}_0 = n$$
$$\begin{pmatrix} R_0 & F_0 & G_0 \\ \tilde{R}_0 & \tilde{F}_0 & \tilde{G}_0 \end{pmatrix} = \begin{pmatrix} \bar{S} & 1 & 0 \\ x^n - 1 & 0 & 1 \end{pmatrix}$$

**while** $d_i \geqslant n - d/2$**:**

$$\mu_i = \mathbf{Coefficient}(R_i, d_i)$$
$$p = d_i - \tilde{d}_i$$

**if** $p \geqslant 0$ **or** $\mu_i = 0$ **then**

$$\begin{pmatrix} R_{i+1} & F_{i+1} & G_{i+1} \\ \tilde{R}_{i+1} & \tilde{F}_{i+1} & \tilde{G}_{i+1} \end{pmatrix} = \begin{pmatrix} 1 & -\mu_i x^p \\ 0 & 1 \end{pmatrix} \begin{pmatrix} R_i & F_i & G_i \\ \tilde{R}_i & \tilde{F}_i & \tilde{G}_i \end{pmatrix}$$
$$d_{i+1} = d_i - 1$$
$$\tilde{d}_{i+1} = \tilde{d}_i$$

**else**

$$\begin{pmatrix} R_{i+1} & F_{i+1} & G_{i+1} \\ \tilde{R}_{i+1} & \tilde{F}_{i+1} & \tilde{G}_{i+1} \end{pmatrix} = \begin{pmatrix} x^{-p} & -\mu_i \\ 1/\mu_i & 0 \end{pmatrix} \begin{pmatrix} R_i & F_i & G_i \\ \tilde{R}_i & \tilde{F}_i & \tilde{G}_i \end{pmatrix}$$
$$d_{i+1} = \tilde{d}_i - 1$$
$$\tilde{d}_{i+1} = d_i$$

**end if**

**end while**

**Return** $F_i, G_i$

**Theorem 5.** *If $t < d/2$ then the algorithm outputs $\Lambda$ and $\Omega$.*

*Proof.* All the conditions in Lemma 4, except item 2, are easily checked using Proposition 1.

For proving item 2 in that proposition notice that, by Proposition 1, $\deg(F_i) = n - \tilde{d}_i$. By looking at the behavior of $d_i$ and $\tilde{d}_i$ we see that at the end of the algorithm, either $\tilde{d}_i = \tilde{d}_0 = n$ or $\tilde{d}_i = d_j$ for some $j < i$, and thus, with $d_j \geqslant n - d/2$. In both cases we get $\deg(F_i) \leqslant d/2$. □

# 5   From the Euclidean to the Berlekamp-Massey algorithm

In this section we will see that the previous algorithm and the Berlekamp-Massey algorithm are essentially the same.

Notice that the only reason to keep the polynomials $R_i$ (and $\tilde{R}_i$) is that we need to compute their leading coefficients (the $\mu_i$'s). We now show that these leading coefficients may be obtained without reference to the polynomials $R_i$. This allows us to compute the $F_i, G_i$ iteratively and dispense with the polynomials $R_i$.

On one hand, the remainder $R_i = F_i \bar{S} - G_i(x^n - 1) = F_i \bar{S} - x^n G_i + G_i$ has degree at most $n - 1$ for all $i \geqslant 0$. This means that all terms of $x^n G_i$ cancel with terms of $F_i \bar{S}$ and that the leading term of $R_i$ must be either a term of $F_i \bar{S}$ or a term of $G_i$ or a sum of a term of $F_i \bar{S}$ and a term of $G_i$.

On the other hand, the algorithm only computes $\text{LC}(R_i)$ while $\deg(R_i) \geqslant n - d/2$. We want to see that in this case the leading term of $R_i$ has degree strictly larger than that of $G_i$. Indeed, one can check that for $i \geqslant 0$, $\deg(G_i) < \deg(F_i)$ and that all $F_i$'s in the algorithm have degree at most $d/2$. So $\deg(G_i) < \deg(F_i) \leqslant d/2 \leqslant n - d/2 \leqslant \deg(R_i)$.